\documentstyle[preprint,aps,prb]{revtex}
\begin{document}
%
%
\newtheorem{lemma}{Lemma}
\newtheorem{theorem}{Theorem}
\newtheorem{corollary}{Corollary}
\newenvironment{duplicate}[2]{\begin{em}\\[0.5\baselineskip]{\bf #1
#2} }{\end{em}\\[0.5\baselineskip]}

\tighten
\preprint{\vbox{\hbox{G\"{o}teborg ITP 94-27}
                \hbox{cond-mat/9411071}
                \hbox{Subm. to Phys. Rev.\ B}}}
\title{Symmetries and Mean-Field Phases of the Extended Hubbard Model }
\author{ Anders B. Eriksson\cite{present_address_Nordita}, Torbj\"{o}rn
Einarsson\cite{present_address_TE}, and Stellan \"{O}stlund} \address{
Institute of Theoretical Physics\\ Chalmers University of Technology
and\\ Gothenburg University\\ S-41296 G\"{o}teborg, Sweden}
\date{\today}
\maketitle
\begin{abstract}
The two-dimensional extended Hubbard model that includes a
nearest-neighbor Heisenberg interaction is studied using a mean-field
theory where quasiparticles are defined by an ${\rm U(8)}$ group of
canonical transformations.  The theory is a generalization of the
ordinary BCS theory, and Balian and Werthamer's theory of $^3{\rm He}$
that permits both broken gauge, spin and sublattice symmetry.  This
allows us to investigate superconductivity, antiferromagnetic order,
charge density waves and, by twisting the spin quantization axis,
spiral antiferromagnetic order in the same theory. Our results for
positive Hubbard $U$ and Heisenberg exchange $J$ suggest that
antiferromagnetic ordering dominates close to half filling, while
spiral states and $d$-wave superconductivity compete when doping is
introduced. For moderate values of $ J $, we find a phase diagram
where a phase transition occurs from an antiferromagnet to a $d$-wave
superconductor as doping is increased.  A narrow region of
$(s+id)$-wave superconductor is found for some values of $J$ and $U$.

\end{abstract}
\draft

\narrowtext

\section{Introduction}
For many years the two-dimensional (2D) Hubbard and ``extended''
Hubbard model have served as simple models of high-$ T_c $ materials
and a paradigm of strongly correlated electrons.  However, despite
many creative attempts to develop approximations to attack the
problem, the 2D Hubbard model has defied a definitive analysis.

In contrast to many of those efforts, the goal in the present paper is
not to try to model the behavior of high-$ T_c $ superconductors or
even to make particularly strong statements about the 2D Hubbard
model. Rather, we wish to make a definitive analysis of the BCS
mean-field theory, an approximation method that has been remarkably
successful in gaining information about simpler model systems with
both broken spin and gauge symmetry.  It is clearly important that we
know what this simple approximation has to say about the Hubbard model
before we can be confident in applying more sophisticated techniques.

Although the program sounds straightforward, the method becomes
surprisingly complicated if one insists on preserving the symmetries
known to exist in the Hubbard model at half filling.  The complication
is due to the exact ${\rm SO(4)} \approx {\rm SU(2)} \times {\rm
SU(2)} $ symmetry\cite{Yang90} precisely at half filling and that this
symmetry mixes many phases that are typically ignored in simpler
calculations.  To do a proper job, we must therefore include all the
order parameters that have been discussed before for this system,
among them spiral
spin-waves,\cite{Shraiman89,Schulz90,Arrigoni91,Dzierzawa92,Sarker91}
N\'eel antiferromagnetism,\cite{Micnas90,Ozaki92} $d$-wave and
$s$-wave
superconductivity\cite{Micnas90,Micnas88a,Micnas88b,Micnas89,Bulka90}
as well as all other phases that are related to these via the symmetry
group.\cite{Ozaki92}

A consequence is that a multitude of possible order parameters must be
retained to provide a self-consistent theory.  Our mean-field theory
systematically enumerates a very large set of representations of the
possible broken symmetries and, among other phases, allows for the
possibility of all types of ordered phases that have been discussed in
previous analyses, including antiferromagnetism, spiral spin-waves and
$d$-wave superconductivity.

The mean-field theory that we employ is based on an $ {\rm U(8)} $ set
of canonical transformations that are analogous to the $ {\rm U(2)} $
canonical transformations discovered by Bogoliubov and Valatin in
their analysis of the broken gauge symmetry in BCS
theory.\cite{Bogoliubov58,Valatin58} Their approach was generalized by
Balian and Werthamer to a set of $ {\rm U(4)} $ transformations in
their study of superfluid $^3{\rm He}$, where also spin rotational
symmetry is broken.\cite{Balian63} We, in turn, extend their analysis
to include broken sublattice symmetry, which extends the relevant
group of canonical transformations to $ {\rm U(8)} $.

The organization of the rest of this paper is the following: After
providing a brief review of related results in
Sec.~\ref{sec:background}, we analyze the symmetry of the Hubbard
model in Sec.~\ref{sec:group-theory}.  The representations of the
${\rm SU(2)}\times{\rm SU(2)}$ symmetry group is made via an $ 8
\times 8 $ Clifford algebra, which provides a generalization of the
Dirac and Pauli matrices used previously in the Nambu formalism
applied in the study of BCS theory and $ ^3{\rm He}
$.\cite{McKenzie90} In Sec.~\ref{sec:ham8x8}, we derive the mean-field
theory for the Hubbard model at half filling and introduce the
point-group symmetry, while we in Sec.~\ref{sec:spiral} extend the
analysis to include the possibility of spiral antiferromagnetic states
that have been proposed as likely phases away from half filling.  We
discuss the reformulation of the self-consistent equations in
Sec.~\ref{sec:low-energy}, and the numerical solution procedure in
Sec.~\ref{sec:numerical}.  Our results are presented in form of phase
diagrams in Sec.~\ref{sec:phase-diagrams}, before the final discussion
in Sec.~\ref{sec:discussion}. Appendices provide additional
mathematical details.

\section{Background}
\label{sec:background}

The Hubbard model is given by the Hamiltonian

\begin{equation}
\label{eqn:Hubbard_ham}
H = H_0 -\mu \hat{N} + H_{\rm Hubb} \,,
\end{equation}
where
\begin{eqnarray}
H_0 &=& -t \sum_{\langle {\bf R},{\bf R}' \rangle,\sigma}
(c^{\dagger}_{{\bf R},\sigma}c^{\phantom{\dagger}}_{{\bf
R}',\sigma}+{\rm h. c.}) \,, \\ \hat{N} &=& \sum_{{\bf R},\sigma}
n_{{\bf R},\sigma}\,, \\ H_{\rm Hubb} &=& U\sum_{{\bf R}} (n_{{\bf
R},\uparrow} -{\textstyle\frac{1}{2}})(n_{{\bf
R},\downarrow}-{\textstyle\frac{1}{2}})\,,
\end{eqnarray}
and where $n_{{\bf R},\sigma}=c^{\dagger}_{{\bf
R},\sigma}c^{\phantom{\dagger}}_{{\bf R},\sigma}$.  The sums in the
hopping terms are summed over all {\em pairs} of nearest neighbors on
the 2D square lattice. The form of the Hubbard term is chosen so as to
make the Hamiltonian particle-hole symmetric, and the system is half
filled (one electron per site) when the chemical potential $\mu$ is
set to zero.

Several extensions to the Hubbard model have been studied.  To have
the possibility of mimicking the behavior of higher order terms which
would be present in a more sophisticated weak-coupling expansion, and
to physically incorporate the effect of spin-spin interactions, we
extend our Hamiltonian by adding the nearest-neighbor Heisenberg
interaction $ H_{\rm Heis} $ given by
\begin{equation}
\label{eqn:Heis_term}
H_{\rm Heis} = J\sum_{\langle {\bf R},{\bf R}' \rangle}{\bf S}_{{\bf
R}} \cdot {\bf S}_{{\bf R}'}\,,
\end{equation}
where ${\bf S}_{{\bf R}}= $ $ \frac{1}{2}(\begin{array}{cc}
c^{\dagger}_{{\bf R},\uparrow} & c^{\dagger}_{{\bf
R},\downarrow}\end{array})\mbox{\boldmath$\sigma$}(\begin{array}{cc}
c^{\phantom{\dagger}}_{{\bf R},\uparrow} & c^{\phantom{\dagger}}_{{\bf
R},\downarrow}\end{array})^T$ and $\mbox{\boldmath$\sigma$}$ is the
vector of Pauli matrices.  The Heisenberg term breaks neither spin nor
pseudospin symmetry, and thus fulfills the most important constraints
on effective higher-order terms in a theory for the pure Hubbard model
at half filling.

In principle, we should also investigate the nearest-neighbor Coulomb
interaction. Although it may be physically important, it violates
pseudospin symmetry which is already broken by the chemical potential
term. Since it adds nothing theoretically fundamental to the model, we
chose to not include it in our analysis.

An appropriate starting point is the pure Hubbard model at half
filling. Here, the model is symmetric under the $ SU(2) \times SU(2) $
group of global spin and ``pseudospin''
rotations.\cite{Yang90,Zhang90a,Zhang91,Zhang90,Ostlund91,Ostlund92,Marston89}
and the phase diagram is rather well understood.

For positive values of $U$, the ground state is expected to be a 2D
antiferromagnetic (AF) $(\pi,\pi)$, or N\'eel, state.  For large
$U/t$, second-order perturbation theory in $t/U$ yields an effective
theory equivalent to the spin-$\frac{1}{2}$ Heisenberg
antiferromagnet. The order parameter is given by the staggered
magnetization, which is a vector which breaks the global ${\rm SU(2)}$
spin and the discrete translational symmetry of the underlying model.

For negative $U$, the groundstate is either an $s$-wave superconductor
(SC) or a $(\pi,\pi)$ charge density wave (CDW). The real and complex
part of the SC order parameter together with the CDW order parameter
form a triplet which transforms as a vector under ${\rm SU(2)}$
pseudospin rotations. The groundstate we call ``mixed'' SC-CDW since
each of these phases are degenerate and related to each other by a
continuous symmetry.

The cases of positive and negative $U$ are directly related via the
``Shiba'' transformation $ Z $.\cite{Shiba} This discrete
transformation has the effect of changing the sign of $ U $ in the
Hubbard model, and also mapping rotations in spin space into rotations
in pseudospin space and vice versa.  Furthermore, a state with broken
pseudospin symmetry for $U<0$ (SC or CDW state), has a direct mapping
to a state with broken spin symmetry (an AF state), and since SC
already appears for an infinitesimal negative value of $U$, AF order
must appear for an infinitesimal positive value. Hence, there is no
Mott transition as a function of $U/t$, and the N\'eel order persists
for all positive values of $U$.

Away from half filling, we are less certain of what phases may occur.
The pseudospin symmetry is broken, with only the $ U(1)$ gauge
subgroup remaining.  Spiral spin-waves, where the antiferromagnetic
order parameter twists with a pitch along a symmetry axis, has been
suggested as a likely candidate.  This phase has been seen in several
theoretical calculations and have been suggested to explain the
incommensurate spin correlations seen
experimentally.\cite{Shraiman89,Schulz90,Arrigoni91,Dzierzawa92,Sarker91}
We study this class of states by imposing a twisted spin quantization
axis in the generalized Hubbard Hamiltonian {\em before} it is
analyzed by the Hartree-Fock-Bogoliubov method.\cite{Arrigoni91}

\section{Group theoretical framework}
\label{sec:group-theory}
\subsection{Using multispinors to represent symmetries of the Hubbard model}

Let us first consider the noninteracting theory $H=H_0$ at $ \mu = 0 $
defined for a square system with $N$ sites, periodic boundary
conditions, and lattice constant 1. This is simply diagonalized as $
H_0 = \sum_{{\bf k}} \epsilon_{{\bf k}} \; c^{\dagger}_{{\bf
k},\sigma} c^{\phantom{\dagger}}_{{\bf k},\sigma} $, where the ${\bf
k}$-sum runs over $N$ points in the first Brillouin zone (BZ).  The
single particle dispersion relation is $ \epsilon_{{\bf k}} = - 2 t
(\cos{k_x}+\cos{k_y})$.  To make manifest the particle-hole symmetry,
we define the vector ${\bf Q}=(\pi,\pi)$ and the operators
\begin{equation}
\begin{array}{ll}
a^{\dagger}_{ {\bf k},\sigma } \equiv c^{\dagger}_{{\bf k},\sigma}\,,
& \; \mbox{when} \hspace{.05in} \; \epsilon_{{\bf k}} < 0\,, \\
b^{\dagger}_{ {\bf k},\sigma } \equiv c^{\dagger}_{{\bf k} + {\bf
Q},\sigma }\,, & \; \mbox{when} \hspace{.05in} \; \epsilon_{{\bf
k}+{\bf Q}} > 0\,.
\end{array}
\end{equation}
In terms of these operators, $H_0$ is written as
\begin{eqnarray}
\label{eqn:H0_ab}
H_0 &=& {\sum_{{\bf k},\sigma}}'' \epsilon_{{\bf k}} (
 a^{\dagger}_{{\bf k} ,\sigma} a^{\phantom{\dagger}}_{{\bf k},\sigma }
 - b^{\dagger}_{{\bf k},\sigma } b^{\phantom{\dagger}}_{{\bf k},\sigma
 } \nonumber\\ &&-a^{\phantom{\dagger}}_{-{\bf
 k},\sigma}a^{\dagger}_{-{\bf k},\sigma} +
 b^{\phantom{\dagger}}_{-{\bf k},\sigma}b^{\dagger}_{-{\bf k},\sigma}
 ) ,
\end{eqnarray}
where the summation denoted by ${\sum}''$ runs over the four times
{\em reduced} Brillouin zone corresponding to $\epsilon_{{\bf k}} < 0
$ and $k_y\ge 0$ (see Fig.~\ref{fig:1BZ}).  Note the special order of
the operators in Eq.~(\ref{eqn:H0_ab}).

We next introduce the ``Shiba'' transformation $Z$ which acts on the
{\em position space} creation and destruction operators $
c^{\dagger}_{{\bf r},\sigma } $ through the canonical transformation
$c^{\dagger}_{{\bf r},\downarrow } \mapsto (-1)^{{\bf r}}
c^{\phantom{\dagger}}_{{\bf r},\downarrow } $, $ c^{\dagger}_{{\bf
r},\uparrow } \mapsto c^{\dagger}_{{\bf r},\uparrow } $, where the
factor $(-1)^{{\bf r}} \equiv e^{ i {\bf Q} \cdot {\bf r} }$ induces a
change of sign on one sublattice. The Shiba transformation is hence a
particle-hole transformation, together with a local change of gauge,
which only acts on the spin-down operators.

In reciprocal space, spin rotations and $ Z $ act naturally on the
8-component multispinor $ {\Psi}^{\phantom{\dagger}}_{{\bf k}} \equiv
$ $ ( a^{\phantom{\dagger}}_{{\bf k},\uparrow},
a^{\phantom{\dagger}}_{{\bf k},\downarrow},
b^{\phantom{\dagger}}_{{\bf k},\uparrow}, b^{\phantom{\dagger}}_{{\bf
k},\downarrow}, a^{\dagger}_{-{\bf k},\uparrow}, a^{\dagger}_{-{\bf
k},\downarrow}, b^{\dagger}_{-{\bf k},\uparrow}, b^{\dagger}_{-{\bf
k},\downarrow} )$ which carries definite momentum $ {\bf k} \bmod{{\bf
Q}}$. In this basis, $Z$ is represented by an idempotent matrix whose
entries are all zero except $ Z_{1,1} = $ $ Z_{3,3} = $ $ Z_{5,5} = $
$ Z_{7,7} = $ $ Z_{2,8} = $ $ Z_{4,6} = $ $ Z_{6,4} = $ $ Z_{8,2} = 1
$, and which acts on ${\Psi}^{\phantom{\dagger}}_{{\bf k}}$ by
matrix-vector multiplication, $ {\Psi}^{\phantom{\dagger}}_{{\bf k}}
\mapsto {\Psi}^{\phantom{\dagger}}_{{\bf k}} Z $.  It is well known
that $ Z $ is an exact symmetry of $ H_0 $ but changes the sign of the
Hubbard term $ U $.  \cite{Ostlund91,Marston89,Shiba,Fradkin91,Lieb68}
It is also simple to verify that the nearest neighbor Heisenberg term
$H_{\rm Heis}$ is invariant under $Z$.

A pseudospin transformation $R'$ is defined as $R'=ZRZ$ where $R$ is
an ordinary global $ SU(2) $ spin rotation.  Since $[H,Z] = 0$ and $
$[H,R]=0, we have $[H,R']=0$, and since $Z$ does not commute with $R$,
we see that the entire symmetry group of $H$ is ${\cal S} =
\mbox{SU(2)}\times\mbox{SU(2)}$.  It follows further that because $
{\Psi}^{\phantom{\dagger}}_{{\bf k}} $ defines a representation of
both $ R $ and $ Z $, it also determines a representation of ${\cal
S}$.

\subsection{A basis spanning the space of Hermitian $8\times 8$ matrices}
\label{ssec:Base_matrices}

To represent the action of spin and pseudospin transformations on the
basis $ {\Psi}^{\phantom{\dagger}}_{{\bf k}} $, we introduce a set of
seven Hermitian $ 8 \times 8 $ matrices $\beta_A$ constructed in
blocks from the ordinary Dirac matrices $\gamma_{\nu}$ as shown in
Table~\ref{tab:gamma_mat}.  By explicit computation, it can be
verified that $ \beta_{A} $ form a Clifford algebra\cite{Miller72},
{\em i.e.}, $ \beta_{A} \beta_{B} + \beta_{B} \beta_{A} = 2 g_{AB} $,
where $ g_{AB} = 0 $ except for $g_{00}=\openone$ and
$g_{nn}=-\openone$, $n=1,\ldots,6$.

The $\beta_A$ matrices have been constructed so that
$(\beta_1,\beta_2,\beta_3)$ transform as a vector under spin rotations
and $(\beta_4,\beta_5,\beta_6)$ transform as a vector under pseudospin
rotations, while $(\beta_4,\beta_5,\beta_6)$ and
$(\beta_1,\beta_2,\beta_3)$ transform as a scalar under spin and
pseudospin rotations, respectively.  Taking multiple products of the
matrices $ \beta_A $, one can construct a complete basis for the
vector space of $ 8 \times 8 $ Hermitian matrices (with real-valued
coefficients).  The Clifford algebra contains four independent
spin/pseudospin scalars, and it follows that there are also four
independent sets of spin and pseudospin vectors and
spin$\otimes$pseudospin tensors, making a total of $4\times(1 + 3 + 3
+ 9) = 64$ basis elements which together span the space of $ 8 \times
8 $ Hermitian matrices.\cite{Ostlund92}

Our goal is to construct a basis that has simple transformation
properties under spin and pseudospin rotations, and with the above
classification scheme in mind, we label the 64 base matrices by $
B_{\mu\nu}^{\kappa}$, where $ 0 \le \mu,\nu,\kappa \le 3$.  The upper
index $\kappa$ enumerates each of the four independent sets of
matrices associated with each of the four scalars.  Transformation
properties under spin rotations are indexed by $ \mu $, where scalars
carry the index $\mu=0$, while $\mu=1,2,3$ represents the components
of a spin vector. Similarly $\nu$ identifies pseudospin scalars and
vectors.  For example, $ B_{01}^2 $ transforms as a scalar under spin
rotations and as the first component of a pseudospin vector.  The
matrices $B_{\mu\nu}^{\kappa}$ are constructed so that they are not
only Hermitian, but also have natural transformation properties under
parity, sublattice exchange and time reversal, and so that the indices
$\mu$ and $\nu$ are simply interchanged under $Z$.

To explicitly construct the basis matrices we introduce the four
scalars.  Three of them are the identity matrix $\openone$, $\beta_0$,
and $\Gamma = i\beta_0\beta_1\beta_2\beta_3 = \left(\begin{array}{cc}
\gamma_5 & 0 \\ 0 & \gamma_5 \end{array}\right) $, where $\gamma_5$
indicates the pseudoscalar Dirac gamma matrix $\gamma_5 = i
\gamma_0\gamma_1\gamma_2\gamma_3$; the fourth scalar matrix is defined
as the product $\beta_0\Gamma$. The four scalars are denoted
$\Upsilon_{\kappa}$ with the following identifications: $\Upsilon_0 =
-\openone$, $\Upsilon_1 = \beta_0$, $\Upsilon_2 = -\Gamma$ and
$\Upsilon_3 = -i\Gamma\beta_0$. To make subsequent formulas simple we
also define $\Omega_0=\hat{\Omega}_0 = \openone$, $\Omega_j =
\beta_j$, and $\hat{\Omega}_j = \beta_{j+3}$, where
$j=1,2,3$. Finally, introducing $\tau_0=1$ and $\tau_j = \sqrt{-1}$,
the basis matrices are defined by
\begin{eqnarray}
B_{\mu,\nu}^{\kappa} & = &
\Upsilon_{\kappa}\Omega_{\mu}\hat{\Omega}_{\nu}\,, \ \ \ \ \mbox{for
$\kappa=0$ or 1} \,,\nonumber \\ B_{\mu,\nu}^{\kappa} & = &
\tau_{\mu}\tau_{\nu}\Upsilon_{\kappa} \Omega_{\mu} \hat{\Omega}_{\nu}
\,,\ \ \ \ \mbox{for $\kappa=2$ or 3}\,,
\end{eqnarray}
where $0\le\mu,\nu\le 3$. The phase factors $\tau_{\mu}$ are chosen so
as to make $B_{\mu\nu}^{\kappa}$ Hermitian.

Letting the subscript ``m'' denote the collection of indices $ \kappa,
\mu, \nu $, we find that the matrices $B_m$ are orthonormal in the
sense that
\begin{mathletters}
\label{eqn:Bas_orthogonality}
\begin{eqnarray}
\mbox{Tr}(B_m B_{m'}) &=& 8 \delta_{m,m'}\,, \\
B_m B_m &=& \openone \,.
\end{eqnarray}
\end{mathletters}
\null It follows that an arbitrary Hermitian
$8\times8$ matrix $M$ can be expanded as
\begin{equation}
M = \sum_m \alpha^m B_m \,,
\end{equation}
where
\begin{equation}
\alpha^m = {\textstyle\frac{1}{8}} \mbox{Tr}( B_m M) \,.
\end{equation}

\section{Writing the mean-field Hamiltonian in terms of $8\times 8$
matrices}
\label{sec:ham8x8}
The interacting Hamiltonian contains not only one-particle but also
two-particle terms.  We use Wick's factorization \cite{Negele88} to
express the expectation values of the two-particle terms as sums of
all products of one-particle terms:

\begin{eqnarray}
\langle O_1O_2O_3O_4 \rangle &=& \langle O_1O_2 \rangle\langle  O_3O_4
\rangle + \langle
O_1O_4 \rangle\langle O_2O_3 \rangle \nonumber \\ && - \langle O_1O_3
\rangle\langle O_2O_4 \rangle\,,
\end{eqnarray}
where $O_i$ denotes any creation or destruction operator.

After transforming $ H + H_{\rm Heis} $ in
Eqs.~(\ref{eqn:Hubbard_ham}) and (\ref{eqn:Heis_term}) to reciprocal
space and then applying the factorization we find
\widetext
\begin{eqnarray}
\langle H \rangle &=& \sum_{{\bf k},\sigma} (\epsilon_{{\bf k}}-\mu)
\langle c^{\dagger}_{{\bf k},\sigma}c^{\phantom{\dagger}}_{{\bf
k},\sigma} \rangle +
\frac{1}{N}
\sum_{\stackrel{{\bf k}_1,{\bf k}_2,{\bf k}_3,{\bf k}_4}{\sigma,\sigma'}}
\bar{\delta}({\bf k}_1-{\bf k}_2+{\bf k}_3-{\bf k}_4)
\left(U\delta_{\sigma,\uparrow}\delta_{\sigma',\downarrow} -
\frac{J}{4}(2\gamma_{{\bf k}_1-{\bf k}_4}+\gamma_{{\bf k}_1 -
{\bf k}_2})\right) \nonumber \\
&&\times \bigg(\langle c^{\dagger}_{{\bf
k}_1,\sigma}c^{\phantom{\dagger}}_{{\bf k}_2,\sigma} \rangle \langle
c^{\dagger}_{{\bf k}_3,\sigma'}c^{\phantom{\dagger}}_{{\bf
k}_4,\sigma'} \rangle + \langle c^{\dagger}_{{\bf
k}_1,\sigma}c^{\phantom{\dagger}}_{{\bf k}_4,\sigma'} \rangle \langle
c^{\phantom{\dagger}}_{{\bf k}_2,\sigma}c^{\dagger}_{{\bf
k}_3,\sigma'} \rangle - \langle c^{\dagger}_{{\bf
k}_1,\sigma}c^{\dagger}_{{\bf k}_3,\sigma'} \rangle \langle
c^{\phantom{\dagger}}_{{\bf k}_2,\sigma}c^{\phantom{\dagger}}_{{\bf
k}_4,\sigma'} \rangle\bigg)\,,
\end{eqnarray}
\narrowtext
\noindent where $N$ is the number of lattice sites,
$\gamma_{{\bf k}}=\cos(k_x)+\cos(k_y)$, and $\bar{\delta}$ indicate a
$\delta$-function modulo the reciprocal lattice.

Let us now consider the one-particle expectation values.  In ordinary
mean-field analysis, all one-particle expectation values that carry
non-zero momentum are assumed to be zero.  Similarly, in our formalism
the natural set of expectation values are related to the elements of
the $8\times 8$ matrix of operators
\begin{equation}
({\Psi}^{\dagger}_{{\bf k}}\otimes{\Psi}^{\phantom{\dagger}}_{{\bf
k}}) \equiv \left( \begin{array}{ccc} a^{\dagger}_{{\bf
k},\uparrow}a^{\phantom{\dagger}}_{{\bf k},\uparrow} &
a^{\dagger}_{{\bf k},\uparrow}a^{\phantom{\dagger}}_{{\bf
k},\downarrow} & \ldots \\ a^{\dagger}_{{\bf
k},\downarrow}a^{\phantom{\dagger}}_{{\bf k},\uparrow} & \ddots & \\
\vdots & & \end{array} \right)
\end{equation}
whose expectation values are all of the form $\langle
c^{\dagger}_{{\bf k}_1,\sigma_1}c^{\phantom{\dagger}}_{{\bf
k}_2,\sigma_2} \rangle$, $\langle c^{\dagger}_{{\bf
k}_1,\sigma_1}c^{\dagger}_{-{\bf k}_2,\sigma_2} \rangle$, and $\langle
c^{\phantom{\dagger}}_{{\bf k}_1,\sigma_1}c^{\phantom{\dagger}}_{-{\bf
k}_2,\sigma_2} \rangle$, where ${\bf k}_1={\bf k}_2$ or ${\bf
k}_1-{\bf k}_2=\pm {\bf Q}$ with ${\bf Q}=(\pi,\pi)$.  Thus we allow
all possible expectation values of operators that carry momentum
$(0,0)$ or $ (\pi,\pi)$, and those that carry spin and/or charge to be
non-zero. Expectation values carrying momentum $(\pi,\pi)$ correspond
to staggered order, net charge represent superconductors, and net spin
is associated with broken spin symmetry.

To make clear the relation between the 64 individual operators in
$({\Psi}^{\dagger}_{{\bf k}}\otimes{\Psi}^{\phantom{\dagger}}_{{\bf
k}})$ and the irreducible representations (irreps) of ${\rm
SU(2)}\times{\rm SU(2)}$ we define a set of 64 operators by the
following linear combinations
\begin{equation}
\label{eqn:alpha_def}
\alpha_{{\bf k}}^m = \frac{1}{8} \mbox{Tr}[B_m
({\Psi}^{\dagger}_{{\bf k}}\otimes{\Psi}^{\phantom{\dagger}}_{{\bf k}})] .
\end{equation}
We then use these operators to rewrite our mean-field Hamiltonian.

We begin by rewriting the quadratic (one-particle) part of the
Hamiltonian in matrix form.  \null From Eq.~(\ref{eqn:H0_ab}) it is
easily seen that $H_0={\sum_{{\bf k}}}'' \epsilon_{{\bf k}}
\mbox{Tr}(B_{00}^1({\Psi}^{\dagger}_{{\bf
k}}\otimes{\Psi}^{\phantom{\dagger}}_{{\bf k}})) = 8{\sum_{{\bf k}}}''
\epsilon_{{\bf k}}(\alpha_{00}^1)_{{\bf k}}$, where the basis matrix
$B_{00}^1$ is diagonal and has diagonal elements $(1,1, -1, -1, -1,
-1, 1, 1)$.  The expectation value of the entire quadratic part of the
Hamiltonian then becomes
\begin{equation}
\label{eqn:mfham_hop}
\langle  H_0-\mu \hat{N} \rangle = 8 {\sum_{{\bf k}}}'' \epsilon_{{\bf k}}
\underbrace{\langle (\alpha_{00}^1)_{{\bf k}} \rangle}_{\mbox{hop}} - \mu
\underbrace{\langle (\alpha_{03}^1)_{{\bf k}} \rangle}_{\mbox{fill}}.
\end{equation}
The interaction (two-particle) part of $\langle H\rangle$ may be
evaluated in a similar manner. To simplify the corresponding
expression, we introduce some shorthand notation. Let
$\langle\alpha_{\mu\nu}^{\kappa}\rangle_{{\bf k}{\bf k}'}^2 \equiv $ $
\langle (\alpha_{\mu\nu}^{\kappa})_{{\bf k}} \rangle \langle
(\alpha_{\mu\nu}^{\kappa})_{{\bf k}'} \rangle$.  With this notation,
the expectation value of the Hubbard interaction
becomes\cite{Ostlund92}
\begin{eqnarray}
\langle  H_{\rm Hubb} \rangle &=& \frac{16U}{N}
{\sum_{{\bf k},{\bf k}'}}''
\underbrace{\langle\alpha_{0i}^3\rangle_{{\bf k}{\bf
k}'}^2}_{\mbox{\small SC-CDW}} -
\underbrace{\langle\alpha_{i0}^3\rangle_{{\bf k}{\bf
k}'}^2}_{\mbox{\small AF}} +
\underbrace{\langle\alpha_{0i}^1\rangle_{{\bf k}{\bf
k}'}^2}_{\mbox{\small SC'-fill}} \nonumber \\ && -
\underbrace{\langle\alpha_{i0}^1\rangle_{{\bf k}{\bf
k}'}^2}_{\mbox{\small FM}}\,,
\label{eqn:mfham_Hubb}
\end{eqnarray}
where the index $i$ is summed from 1 to 3.  The terms representing the
mixed superconducting and charge density wave (SC-CDW),
antiferromagnetic (AF) and ferromagnetic (FM) ordering are
underbraced. The third (``z'') component of the SC'-fill term is
recognized as the filling while the $x$ and $y$ components represent
the real and imaginary part of a kind of staggered superconductor
(SC'). This term together with the first are the analogues of the FM
and AF states under the change of sign of $U$. We find numerically
that the order parameter for SC' is always zero for the interaction
parameters that we consider.

To make a more systematic investigation of all order parameters, we
note that the Hubbard model on the 2D square lattice has the point
group symmetry $ C_{4v} $, which has four 1D irreps ($A_1$, $A_2$,
$B_1$ and $B_2$) and one 2D irrep ($E$) (see App.~\ref{app:C4} ).  The
total symmetry is hence $ C_{4v} \times {\rm SU(2)} \times {\rm SU(2)}
$, and it is possible to perform a complete classification of the
order parameters.\cite{Ozaki92}

Since the Hubbard interaction is on-site, the interaction coefficients
are independent of ${\bf k}$ and all the prefactors of the Hubbard
term belong to the $A_1$ irrep.  Another basis function for this
representation is $\gamma_{{\bf k}} = \cos(k_x)+\cos(k_y)$, which
occurs in the hopping and Heisenberg terms and in ``extended
$s$-wave'' superconductivity.  The Heisenberg term has also components
in the $B_1$ and the $E$ representations.  The $B_1$ terms (odd under
$k_x \leftrightarrow k_y$) have the ${\bf k}$-dependence $\eta_{{\bf
k}}= \cos(k_x)-\cos(k_y)$ and the two basis-functions of the $E$
representation (odd under parity, ${\bf k}\to-{\bf k}$) are
$\zeta_{k_x} = \sin(k_x)$ and $\zeta_{k_y} = \sin(k_y)$.  The
expectation value of the Heisenberg term reads, in its whole glory,
\widetext
\begin{eqnarray}
\langle  H_{\rm Heis} \rangle &=& \frac{16J}{N}
{\sum_{{\bf k},{\bf k}'}}'' 2
\underbrace{\langle\alpha_{i0}^1\rangle_{{\bf k}{\bf k}'}^2}_{
\mbox{\small FM}} - 2 \underbrace{\langle\alpha_{i0}^3\rangle_{{\bf
k}{\bf k}'}^2}_{\mbox{\small AF}}
\nonumber \\ &&
+ {\textstyle\frac{1}{4}}(\gamma_{{\bf k}}\gamma_{{\bf k}'} + \eta_{{\bf
k}}\eta_{{\bf k}'}) ( -3 \underbrace{\langle\alpha_{00}^1\rangle_{{\bf
k}{\bf k}'}^2}_{ \mbox{\small hop}} - 3
\underbrace{\langle\alpha_{0i}^2\rangle_{{\bf k}{\bf k}'}^2}_{\mbox{
\small SC-CDW}} + \underbrace{\langle\alpha_{i0}^2\rangle_{{\bf k}{\bf
k}'}^2}_{ \mbox{\small spin nem.}} + \langle\alpha_{ij}^1\rangle_{{\bf
k}{\bf k}'}^2 ) \nonumber \\ && +
{\textstyle\frac{1}{2}}(\zeta_{k_x}\zeta_{k_x'} +
\zeta_{k_y}\zeta_{k_y'}) ( -3\langle\alpha_{00}^3\rangle_{{\bf k}{\bf
k}'}^2 -3\langle\alpha_{0i}^0\rangle_{{\bf k}{\bf k}'}^2 +
\langle\alpha_{i0}^0\rangle_{{\bf k}{\bf k}'}^2 +
\underbrace{\langle\alpha_{ij}^3\rangle_{{\bf k}{\bf k}'}^2
}_{\mbox{\small $p$-wave}}) .
\label{eqn:mfham_Heis}
\end{eqnarray}
\narrowtext
\noindent From this form we see that the Heisenberg term renormalizes the
hopping and also affects the antiferromagnetic and ferromagnetic
ordering. Apart from these terms, the most interesting term is the
$\alpha_{0i}^2$-term representing superconducting and CDW
ordering. This term is split into two irreps, $A_1$ and $B_1$
corresponding to $\gamma_{{\bf k}}$ and $\eta_{{\bf k}}$. The $A_1$
part gives rise to ``extended $s$-wave superconductivity'' in certain
regions of parameter space.  The $B_1$ part introduces $d$-wave
superconductivity, and the $z$-component of this part is recognized as
an orbital antiferromagnet.  We also recognize the $B_1$
representation of the $\alpha_{i0}^2$ term, which represents a spin
nematic state with the order parameter $\langle c^{\dagger}_{{\bf
k},\alpha}c^{\phantom{\dagger}}_{{\bf k},\beta}\rangle = i \eta_{{\bf
k}} \mbox{\boldmath$\sigma$}_{\alpha\beta}\cdot{\bf d}$, where ${\bf
d}$ is a real vector.  Finally, there is a term representing $p$-wave
superconductivity and since it is odd under parity it belongs to the
two-dimensional $E$ representation. Some of the other terms may have
been discussed in the literature but, since we have found them not to
be energetically favored, we do not discuss them further.

In addition, the full Hamiltonian gives a term that shifts the
chemical potential, and the Hubbard and chemical-potential terms
produce constants that shift the energy. These terms have no
significance in our analysis and they are neglected in the following.

In collecting all these terms, it is useful to write the expectation
value of the Hamiltonian in the following generic form:
\begin{equation}
\label{eqn:generic_ham}
\langle H \rangle = {\sum_{l,m,{\bf k}}}''a_{l,m} \beta_{{\bf k}}^l
\langle \alpha_{{\bf k}}^m \rangle +
{\sum_{l,m,{\bf k},{\bf k}'}}'' b_{l,m} \beta_{{\bf k}}^l\beta_{{\bf
k}'}^l \langle \alpha_{{\bf k}}^m \rangle \langle \alpha_{{\bf k}'}^m
\rangle \,,
\end{equation}
where $a_{l,m}$ and $b_{l,m}$ are coefficients of the linear and
quadratic terms, respectively, and $\beta_{{\bf k}}^l=1,\gamma_{{\bf
k}},\eta_{{\bf k}},\ldots$ are trigonometric prefactors from various
irreps of the point group $C_{4v}$.  The index ``l'' labels the irreps
of $ C_{4v} $, while the index ``m'' labels the irreps of $ {\rm
SU(2)} \times {\rm SU(2)} $.

\section{Incorporating the spiral phases in the Hamiltonian}
\label{sec:spiral}
The most important unanswered question is what happens when the
electronic density deviates from half filling.  A widely discussed
scenario is that the antiferromagnetic state adjusts to incorporate
the excess electrons and changes into a spiral phase --- an
antiferromagnetic phase whose order parameter shows a spiral spatial
distribution.  Given that the spiral antiferromagnet is the
energetically preferred phase in some other mean-field calculations,
we must clearly incorporate this scenario in our analysis.
Unfortunately, spiral states cannot be directly generated by the
Bogoliubov transformations we have used this far.  We therefore
generalize our method by imposing a spiral twist of the quantization
axis on the Hamiltonian {\em before} the mean-field approximation is
applied.

To describe any spiral we need a 2D vector ${\bf q}$ in the plane,
which controls the direction and pitch of the spiral spin wave, and a
3D unit vector ${\bf \Omega}$ which defines the axis around which the
spin is twisted.  The spin-twisting canonical transformation is then
generated by the operator
\begin{equation}
\left(\begin{array}{cc} c^{\dagger}_{{\bf r},\uparrow} &
c^{\dagger}_{{\bf r},\downarrow}
\end{array} \right) \exp[i({\bf q}\cdot{\bf r})({\bf \Omega} \cdot
\mbox{\boldmath$\sigma$})] \left(\begin{array}{cc}
c^{\phantom{\dagger}}_{{\bf r},\uparrow} \\
c^{\phantom{\dagger}}_{{\bf r},\downarrow} \end{array} \right)\,,
\end{equation}
where $\mbox{\boldmath$\sigma$}$ is the vector of Pauli matrices. This
canonical transformation is easily written as a unitary transformation
of the creation/annihilation operators,
\begin{equation}
\left(\begin{array}{c}c^{\dagger}_{{\bf r},\uparrow}\\
c^{\dagger}_{{\bf r},\downarrow}
\end{array}\right) \rightarrow [ \cos({\bf q}\cdot{\bf r}) \openone + i
\sin({\bf q}\cdot{\bf r}) {\bf \Omega}\cdot\mbox{\boldmath$\sigma$} ]
\left(\begin{array}{c} c^{\dagger}_{{\bf r},\uparrow}\\
c^{\dagger}_{{\bf r},\downarrow}
\end{array}\right)  .
\end{equation}
Since the Hamiltonian is invariant under global spin rotations the
direction of ${\bf \Omega}$ can be arbitrarily set to be along
$\hat{z}$.

When applying this spiral spin transformation to $H$, the Hubbard
interaction is invariant since it is spin-rotation invariant and
on-site, but the hopping term transforms into $H_0^{{\bf q}} =
\sum_{{\bf k}} \epsilon_{{\bf k}}(n_{{\bf k}+{\bf q},\uparrow} +
n_{{\bf k}-{\bf q},\downarrow})$, which, using our operator
definitions, is written as
\begin{eqnarray}
\langle H_{0}^q \rangle &=& -2t{\sum_{{\bf k},n=x,y}}''
8[\cos(q_{n})\cos(k_{n}) \langle (\alpha_{00}^1)_{{\bf k}} \rangle
\nonumber \\ && + \sin(q_{n})\sin(k_{n})\langle (\alpha_{30}^0)_{{\bf
k}} \rangle]\,.
\end{eqnarray}
The Heisenberg term transforms into $H_{\rm Heis}^{{\bf q}} =
J\sum_{\langle {\bf R},{\bf R}' \rangle} \vec{S}_{{\bf
R}}\cdot\vec{S}_{{\bf R}'} \cos[2{\bf q}\cdot({\bf R}-{\bf R}')] + 2
S_{{\bf R}}^z S_{{\bf R}'}^z \sin^2[{\bf q}\cdot({\bf R}-{\bf R}')]$,
where $S_{{\bf R}}^z$ denotes the $z$-component of the local spin
operator at site ${\bf R}$.  Using the same notation as in
Eqs.~(\ref{eqn:mfham_Hubb}) and (\ref{eqn:mfham_Heis}) with the
extension that $\iota$ is an index that is summed over 1 and 2 only,
the Heisenberg expectation value is
\widetext
\begin{eqnarray}
\langle H_{\rm Heis}^{\bf q} \rangle &=& \frac{16J}{N}
{\sum_{\stackrel{{\bf k},{\bf k}'}{n=x,y}}}''
\Bigg[\langle\alpha^{1}_{30}\rangle_{{\bf k}{\bf
k}'}^2-\langle\alpha^{3}_{30}\rangle_{{\bf k}{\bf k}'}^2 + \cos(q_{n})
[\langle\alpha^{1}_{\iota0}\rangle_{{\bf k}{\bf k}'}^2 -
\langle\alpha^{3}_{\iota0}\rangle_{{\bf k}{\bf k}'}^2]
\nonumber \\ &&
+ \frac{1}{2}\cos(k_{n})\cos(k_{n}')
\bigg\{-[1+2\cos(2q_{n})] [\langle\alpha^{1}_{00}\rangle_{{\bf k}{\bf
k}'}^2+\langle\alpha^{2}_{0i}\rangle_{{\bf k}{\bf k}'}^2] +
\langle\alpha^{2}_{\iota0}\rangle_{{\bf k}{\bf
k}'}^2+\langle\alpha^{1}_{\iota i}\rangle_{{\bf k}{\bf
k}'}^2
\nonumber \\ &&
+[2\cos(2q_{n})-1][\langle\alpha^{2}_{30}\rangle_{{\bf k}{\bf
k}'}^2+\langle\alpha^{1}_{3i}\rangle_{{\bf k}{\bf k}'}^2]\bigg\}
\nonumber \\ &&+\sin(k_{n})\sin(k_{n}')\bigg\{ -[1+2\cos(2q_{n})]
[\langle\alpha^{3}_{00}\rangle_{{\bf k}{\bf k}'}^2 +
\langle\alpha^{0}_{0i}\rangle_{{\bf k}{\bf k}'}^2] +
\langle\alpha^{0}_{\iota 0}\rangle_{{\bf k}{\bf k}'}^2 +
\langle\alpha^{3}_{\iota i}\rangle_{{\bf k}{\bf k}'}^2 \nonumber \\ &&
+ [2\cos(2q_{n}]-1][\langle\alpha^{0}_{30}\rangle_{{\bf k}{\bf k}'}^2
+\langle\alpha^{3}_{3i}\rangle_{{\bf k}{\bf k}'}^2] \bigg\} \Bigg]\,.
\end{eqnarray}
\narrowtext
The large number of terms in this expression is due to the broken
spin-rotational and point-group symmetries.

\section{Solving for the state of lowest free energy}
\label{sec:low-energy}
\subsection{Self-consistent equations at finite temperature}

In the spirit of standard BCS theory we introduce the reduced
Hamiltonian
\begin{equation}
\label{eqn:reduced_ham}
H = {\sum_{l,m,{\bf k}}}''a_{l,m} \beta_{{\bf k}}^l \alpha_{{\bf k}}^m
+ {\sum_{l,m,{\bf k},{\bf k}'}}'' b_{l,m} \beta_{{\bf k}}^l\beta_{{\bf
k}'}^l \alpha_{{\bf k}}^m\alpha_{{\bf k}'}^m \,,
\end{equation}
which has the generic expectation value given in
Eq.~(\ref{eqn:generic_ham}).  We also define the mean-field order
parameters (gap functions)
\begin{equation} \label{eqn:Delta_def}
\Delta_{l,m} ={\sum_{{\bf k}}}''\beta_{{\bf k}}^l\langle
\alpha_{{\bf k}}^m \rangle\,.
\end{equation}
Using the assumption that the fluctuations in the operators $
\alpha_{{\bf k}}^m $ from their mean-field values are small, we
substitute
\begin{equation}
\alpha_{{\bf k}}^m  = ( \alpha_{{\bf k}}^m - \langle
\alpha_{{\bf k}}^m \rangle) + \langle\alpha_{{\bf k}}^m\rangle
\end{equation}
into $ H $ in Eq.~(\ref{eqn:reduced_ham}), drop terms quadratic in $ (
\alpha_{{\bf k}}^m - \langle\alpha_{{\bf k}}^m\rangle ) $ and find the
mean-field Hamiltonian $H_{\rm mf}$,
\begin{equation}
\label{eqn:generic_mf_ham}
H_{\rm mf} = {\sum_{l,m,{\bf k}}}'' (a_{l,m} + 2 b_{l,m} \Delta_{l,m}
)\beta_{{\bf k}}^l \alpha_{{\bf k}}^m - \sum_{l,m} \Delta_{l,m}^2 .
\end{equation}
Aside from a constant that is unimportant in this discussion, $H_{\rm
mf}$ can be recast in the form
\begin{eqnarray}
H_{\rm mf} & = & {\textstyle\frac{1}{8}} { \sum_{{\bf k}, \alpha \beta
 }}'' h_{\beta \alpha}({\bf k}) ( \Psi^{\dagger}_{{\bf k}} )^{\alpha}
 ( \Psi_{{\bf k}} )^{\beta} \nonumber \\ & = & {\textstyle\frac{1}{8}}
 {\sum_{{\bf k}}}'' \mbox{Tr}[h({\bf k}) ({\Psi}^{\dagger}_{{\bf
 k}}\otimes{\Psi}^{\phantom{\dagger}}_{{\bf k}})]\,,
\end{eqnarray}
where
\begin{equation} \label{eqn:hdef}
h({\bf k}) = \sum_{l,m}(a_{l,m} + 2 b_{l,m} \Delta_{l,m} )\beta_{{\bf
k}}^l B_{m}\,.
\end{equation}
Introducing the matrix of expectation values
\begin{equation} \label{eq:fk_def}
f_{{\bf k}} \equiv \langle ({\Psi}^{\dagger}_{{\bf
k}}\otimes{\Psi}^{\phantom{\dagger}}_{{\bf k}}) \rangle\,,
\end{equation}
it follows from Eqs.~(\ref{eqn:alpha_def}) and (\ref{eqn:Delta_def})
that
\begin{equation} \label{eqn:self_consist_eqn}
\Delta_{l,m} =  {\textstyle\frac{1}{8}} {{\sum_{{\bf k}}}''}
\beta_{{\bf k}}^l
         \mbox{Tr} \left( B_m f_{{\bf k}} \right) .
\end{equation}
To evaluate $f_{{\bf k}}$ we note that the mean-field Hamiltonian is
bilinear in ${\Psi}^{\dagger}_{{\bf k}}$ and can be diagonalized,
\begin{equation}
H_{\rm mf} = {\sum_{\alpha, {\bf k} }}'' \epsilon_{\alpha}( {\bf k} )
({\chi}^{\dagger}_{{\bf k}})^{\alpha}
({\chi}^{\phantom{\dagger}}_{{\bf k}})^{\alpha} \,,
\end{equation}
by the canonical transformation
\begin{equation}
{\chi}^{\dagger}_{{\bf k}} = U^{\phantom{\dagger}}_{{\bf k}}
{\Psi}^{\dagger}_{{\bf k}}\,,
\end{equation}
where $ U_{{\bf k}} $ is an ${\rm U(8)}$ matrix.  Standard arguments
from statistical mechanics give in the diagonal case
\begin{equation} \label{eqn:sc_Psimat}
\langle ({\chi}^{\dagger}_{{\bf k}}\otimes{\chi}^{\phantom{
\dagger}}_{{\bf k}})_{\alpha, \gamma}  \rangle =
	( 1 + e^{\beta \epsilon_{\alpha} ({\bf k}) } )^{-1}
        \delta_{\alpha\gamma} \,,
\end{equation}
($\beta$ is the inverse temperature, $\beta=1/k_B T$) which, when
transformed back to the operators $\Psi_{{\bf k}}$, results in
\begin{equation} \label{eqn:f_def}
f_{{\bf k}} = \langle ({\Psi}^{\dagger}_{{\bf
k}}\otimes{\Psi}^{\phantom{\dagger}}_{{\bf k}}) \rangle = ( 1 + e^{
\beta h({\bf k}) } )^{-1}\,.
\end{equation}
Equations (\ref{eqn:hdef}), (\ref{eqn:self_consist_eqn}), and
(\ref{eqn:f_def}) constitute the self-consistent equations to be
solved for $ \Delta_{l,m} $.  The solutions extremize the free energy
$F$,
\begin{equation} \label{eqn:free_energy}
F = \langle H \rangle - TS \, ,
\end{equation}
where $\langle H \rangle$ now includes the chemical potential, and
where $S$ is the entropy.  If more than one solution is found, then
the one with the minimal value of $F$ is the physical one. To
calculate $F$, both terms in Eq.~(\ref{eqn:free_energy}) must be
explicitly evaluated.  The evaluation of $\langle H \rangle$ is
straightforward --- using Eq.~(\ref{eqn:generic_ham}) we see that it
is equal to
\begin{equation} \label{eqn:Feq1}
\langle H \rangle = \sum_{l,m}
( a_{l,m} \Delta_{l,m} + b_{l,m} \Delta_{l,m}^2 ) \,.
\end{equation}
The entropy is in turn given by
\begin{equation} \label{eqn:Feq2}
S = -k_B {\sum_{{\bf k}}}'' \mbox{Tr}[f_{{\bf k}} \ln{ f_{{\bf k}} } +
                        ( 1 - f_{{\bf k}} ) \ln{ ( 1 - f_{{\bf k}} )
                        }]\,,
\end{equation}
where $ f_{{\bf k}} $ is the matrix of expectation values defined in
Eq.~(\ref{eqn:f_def}).

\subsection{Zero temperature}

At zero temperature, the variational state gives an approximate ground
state $| G \rangle$ and the thermal expectation values evolve into
expectation values with respect to this ground state {\em i.e.},
$\langle ({\Psi}^{\dagger}_{{\bf
k}}\otimes{\Psi}^{\phantom{\dagger}}_{{\bf k}}) \rangle$ is replaced
by $\langle G|{\Psi}^{\dagger}_{{\bf
k}}\otimes{\Psi}^{\phantom{\dagger}}_{{\bf k}}|G\rangle$.  The
zero-temperature self-consistent equations are obtained from
Eq.~(\ref{eqn:self_consist_eqn}) in the limit $\beta\to \infty$.

Combining the facts that the ground state is the vacuum state for the
quasiparticles and that the first (last) four elements of
${\chi}^{\phantom{\dagger}}_{{\bf k}}$ are annihilation (creation)
operators, we can identify the diagonal matrix $g = \langle
G|({\chi}^{\dagger}_{{\bf k}}\otimes{\chi}^{\phantom{\dagger}}_{{\bf
k}})|G\rangle$.  Its diagonal entries are $(0,0,0,0,1,1,1,1)$, and
from Eq.~(\ref{eqn:sc_Psimat}) with $\beta\to\infty$ it follows that
four of the eigenvalues of $ h({\bf k}) $ must be negative and the
rest positive.  We then solve for the unitary matrices $U_{{\bf k}}$
that diagonalize the $h({\bf k})$'s defined in Eq.~(\ref{eqn:hdef}),
in such a way that the eigenvalues in the diagonal matrices $D_{{\bf
k}}$ are in descending order.  The self-consistent equations to be
iterated are then Eq.~(\ref{eqn:hdef}) and
\begin{mathletters}
\label{eqn:T0_self_cons}
\begin{eqnarray}
\Delta_{m}  &=& \frac{1}{8}{\sum_{{\bf k}}}''\beta_{{\bf k}}^m\mbox{Tr}(
B_mU_{{\bf k}}^{\dagger} g U_{{\bf k}}^{\phantom{\dagger}}) \,,\\
D_{{\bf k}} &=& U_{{\bf k}}^{\phantom{\dagger}}h({\bf k})U_{{\bf
k}}^{\dagger} \,.
\end{eqnarray}
\end{mathletters}

\subsection{An alternative set of equations for the ground
state at half filling} The self-consistent equations must be solved
numerically.  Although the equations are formally simple, it is quite
challenging to numerically carry out a search of the solution space.
We therefore present an alternative method to find the ground state,
and use it to derive a theorem of stability of the superconducting and
the antiferromagnetic solutions at half filling for some regimes of
$U$ and $J$.

Na\"{\i}vely, the expectation value of the Hamiltonian,
Eq.~(\ref{eqn:generic_ham}), is a simple quadratic form and one should
just find its minimum. However, since the expectation values $\langle
\alpha_{{\bf k}}^m \rangle$ are constrained by the fact that they
represent expectation values of fermion operators, the terms cannot be
independently varied, and there are constraints on the set of
$\langle\alpha_{{\bf k}}^m\rangle$.  These restrictions were
automatically satisfied in the previous analysis, since $\alpha_{{\bf
k}}^m$ was explicitly computed through a canonical transformation.
Another way to proceed is to try to find a constrained quadratic
minimum directly without first calculating a canonical transformation.

To express the fermionic constraints, we define the following matrix
\begin{equation}
\label{eqn:Ap_def}
A_{{\bf k}} = \sum_{\stackrel{m\neq m'}{m,m'\neq 0}}
\langle\alpha_{{\bf k}}^m\rangle \langle\alpha_{{\bf k}}^{m'}\rangle
\{B_m,B_{m'}\}\,,
\end{equation}
where $\{A,B\}$ denotes the anticommutator $ \{A,B\} \equiv AB+BA$.
The matrix $B_0 \equiv B_{00}^0 = -\openone$ that is excluded from the
sum is the only basis matrix with nonzero trace.  First we state the
lemma that reformulates the problem of minimizing the energy.
\begin{lemma}
\label{lem:T0}
The restrictions on the expectation values $\langle\alpha_{{\bf
k}}^m\rangle$ for the variational solutions of
Eq.~(\ref{eqn:generic_ham}) are: $\sum_{m} \langle\alpha_{{\bf
k}}^m\rangle^2 = \frac{1}{2}$, $\langle(\alpha_{00}^0)_{{\bf
k}}\rangle=-\frac{1}{2}$, and $A_{{\bf k}}=0 \; \forall\; {\bf k}$.
\end{lemma}
This lemma is proved in App.~\ref{app:proofsT0}.  The importance of
the lemma is that it shows that the minimization problem of
Eq.~(\ref{eqn:generic_ham}) is a quadratic minimum subject to
quadratic constraints.  This enables us to search for minima of the
{\em unconstrained} problem, which, if they are found to satisfy the
constraints, must also be minima of the constrained problem. A class
of such solutions are introduced in the following theorem (proved in
App.~\ref{app:proofsT0}) and corollary.
\begin{theorem}
\label{th:T0}
Consider the Hamiltonian in Eq.~(\ref{eqn:generic_ham}) with fixed
coefficients $a_{l,m}$ and $b_{l,m}$. Define $\cal N$ as the subset of
purely quadratic and irreducible terms, ${\cal N} = \{ m: a_{l,m}=0 \,
\forall\, l\ , b_{l,m}=b_m \delta_{l,l(m)} \}$, where $l(m)$ is a
function which attaches one single $l$ to each $m$.  Assume there
exists an $\tilde{m}\in{\cal N}$, such that ($b_{\tilde{m}}<0$ and
$b_{\tilde{m}}|\beta_{{\bf k}}^{l(\tilde{m})} \beta_{{\bf
k}'}^{l(\tilde{m})}| < b_m|\beta_{{\bf k}}^{l(m)} \beta_{{\bf
k}'}^{l(m)}| \, \forall\, {\bf k},{\bf k}'$, $\forall\, m \in {\cal
N}\setminus\{\tilde{m}\}$).  If ($\langle\alpha_{{\bf k}}^m\rangle=0\,
\forall\, {\bf k},\;\forall\, m\in{\cal N}\setminus\tilde{m}$) is a
sufficient condition for ($A_{{\bf k}}=0 \,\forall\, {\bf k}$), then
the same $\langle\alpha\rangle$'s will be zero also in the minimizing
solution of the constrained problem.
\end{theorem}
In the case of half-filling the following corollary now follows for
the two important cases of antiferromagnetic and $s$-wave
superconducting ordering.
\begin{corollary}
\label{corr:T0}
The state of lowest energy at half filling ($\mu=0$) for the
Hamiltonian ($H=H_0+H_{\rm Hubb}+H_{\rm Heis}$) in
Eqs.~(\ref{eqn:mfham_hop}), (\ref{eqn:mfham_Hubb}) and
(\ref{eqn:mfham_Heis}) is AF ($\langle\alpha_{i0}^2\rangle\neq 0$) if
$0<J<U$, and $s$-wave SC-CDW ($\langle\alpha_{0i}^2\rangle \neq 0$) if
$U<J<-U/3$.
\end{corollary}
The theorem follows by making two observations. First, the possible
low-energy states follow from Theorem~\ref{th:T0} by inspection of the
prefactors of the quadratic terms in the Hamiltonian, and by verifying
that $A_{{\bf k}}=0$ if all $\langle\alpha\rangle$'s are zero except
the hopping $\langle\alpha_{00}^1\rangle$ and either
$\langle\alpha_{i0}^2\rangle$ or $\langle\alpha_{0i}^2\rangle$.
Secondly, we observe that the problem that results by setting all
other order parameters to zero is analogous to the ordinary $s$-wave
SC case, where it is well known that any attractive interaction
results in a finite order parameter.

We further note that there are two degenerate superconductivity
solutions if the constraints are disregarded. One is the ordinary
$s$-wave superconductor, and the other is the staggered
superconductor. However, the latter is ruled out by the fact that the
constraint $A_{{\bf k}}=0$ is not fulfilled. Apart from predicting
these low energy states, Theorem~\ref{th:T0} also proves the stability
of these phases with respect to small perturbations to the
Hamiltonian. Since the ferromagnetic state has $A_{{\bf k}}\neq 0$,
ferromagnetic ordering and nonzero hopping cannot be present
simultaneously, at least not at the same location in ${\bf
k}$-space. The predictions of the Corollary are illustrated in
Fig.~\ref{fig:ph_corr}.

\section{Numerical methods}
\label{sec:numerical}
\narrowtext
To find the phase at a $(U, J,\mu,T)$-point in a phase diagram, we
choose an initial value of the pitch vector ${\bf q}$, solve the
self-consistent equations Eqs.~(\ref{eqn:hdef}),
(\ref{eqn:self_consist_eqn}) and (\ref{eqn:f_def}) numerically, and
calculate $F({\bf q})$ using Eqs.~(\ref{eqn:Feq1}) and
(\ref{eqn:Feq2}).

The pitch vector $ {\bf q} $ is then varied to find which $ {\bf q}$
gives the solution to the self-consistent equations with the lowest
value of $F({\bf q})$.  The non-zero mean-field order parameters
$\Delta_{l,m}$ defines, together with ${\bf q}$, the particular phase
for the $(U,J,\mu,T)$-point in parameter space.

Complete phase diagrams are obtained by the following two steps:
\begin{itemize}

\item The parameters $(U,J,\mu,T,{\bf q})$ are swept,
with an initial set of $\Delta_m$'s generated at random and the
self-consistent equations are iterated until a fixed point is
reached. This gives us a rough picture of the states that are present
in the phase diagram.

\item  The accuracy of the boundaries between the phases in the
phase diagram is improved. Here the self-consistent equations are
solved using Broyden's, method\cite{Press92} which is often more
efficient than the previous iterative method.

\end{itemize}

To find $ \Delta_{l,m} $ for a particular value of ${\bf q}$, we cover
the reduced Brillouin zone by a discrete lattice.  Care has to be
taken not to break any of the symmetries of the problem.
Fig.~\ref{fig:1BZ} shows how a 32-point lattice is laid out.  The most
time-consuming numerical step is to diagonalize the $8\times 8$ matrix
in the argument of the exponential function in Eq.~(\ref{eqn:f_def})
at every ${\bf k}$-point; this must be done for each iteration.

Choosing random order parameters as initial conditions for recursion
is useful when there is no a priori information about the expected
solutions of the self-consistent equations.  A complication of this
method, however, is that the iteration tends to fall into cycles.  We
cured this by including a tail of exponentially damped previous
iterates at each step.  Sometimes, the procedure still did not
converge, and several initial points must be used before a fixed point
was found.

To obtain a complete $\mu$--$T$ phase diagram for fixed $U$ and $J$,
we cover the $(\mu,T,{\bf q})$ space with roughly 1500 points, and
repeat the iterative procedure 10 times.  On an ordinary workstation
it takes of the order of a week of CPU time to trace out the phase
diagram using 98 points in the reduced Brillouin zone and solving the
self-consistent equations to an accuracy of 1 \%.  Of course,
solutions could be missed by chance since we use random initial
guesses, and phases occurring in narrow regions of the phase space
could be missed since the parameter space is not covered with a fine
enough mesh.

After the different phases have been identified, we obtain more
accurate solutions of the self-consistent equations using Broyden's
method.  This method cannot be used from the beginning since the
initial guess has to be close to the final answer for the method to
converge.  The method is also slow if the number of order parameters
is very large.  Here, we therefore eliminate from the Hamiltonian all
order parameters that are known to be zero in the corresponding
regions.

If the state of lowest free energy has ${\bf q}=0$ the minimizing
solution can be obtained directly by Broyden's method, and in this
case the phase boundaries are located to high accuracy; generally by
using 800 points in the reduced Brillouin zone.

If ${\bf q}$ is non-zero, there is the extra problem of minimizing the
free energy with respect to ${\bf q}$.  We do this by using Broyden's
method for solving the self-consistent equations and extending it with
a simple numerical algorithm that also minimizes in ${\bf q}$ using
Brent's method.\cite{Press92} To obtain results within reasonable
computing time, most of the spiral spin-wave calculations have been
performed using a $392$-point lattice in the reduced Brillouin zone.

In order to allow for the most general spiral solutions, no
restrictions should be imposed on the spiral spin-wave parameter ${\bf
q}$, but that would make the problem numerically unmanageable.
Instead we have focused on the question of whether the low-energy
state is a spiral spin-wave or not. Assuming the spiral spin-wave not
to break the lattice symmetries completely, the quantization axis
${\bf q}$ could be twisted either in the $(1,1)$ or $(1,0)$
direction. We further concentrated on the latter case case, ${\bf q} =
(q,0)$, since it turned out to give a slightly lower free energy than
the diagonal twist in some regions of the phase diagram.

\narrowtext

\section{Phase diagrams}
\label{sec:phase-diagrams}
In order to present a set of complete phase diagrams for the extended
Hubbard model, we would have to probe all combinations of values of
the four parameters $U$, $J$, filling (or $\mu$) and $T$.  This is an
unfeasible task, and we restrict ourselves to certain cross sections
that we hope to capture generic behavior.  Due to the particle-hole
symmetry as $\mu \to - \mu$, we restrict our phase diagrams to hole
doping ($\mu < 0$).  We can further set $t=1$ without loss of
generality.

\subsection{Zero temperature and half filling}
The most fundamental cross section is $(T = 0,\mu = 0)$ corresponding
to the ground state of the half-filled extended Hubbard model in
Fig.~\ref{fig:pd_T0_mu0}.  In this case, when $U$ and $J$ are both
positive, as well as for a region of negative $U$, the
antiferromagnetic (AF) state forms a numerically very stable solution.
This is consistent with Corollary~\ref{corr:T0}.  The corollary
further indicates a region (SC-CDW) of degenerate $s$-wave SC and
charge-density-wave state, which is confirmed by the numerical
simulations for negative $U$ and intermediate $J$.

An interesting feature occurring in a region outside the validity of
the assumptions of Corollary~\ref{corr:T0}, but which is numerically
very robust, is the tongue of $d$-wave SC-CDW, $(s+id)$-wave SC-CDW and
($d$-wave SC-CDW)+AF between the AF and the $s$-wave states. The
$d$-wave order parameter is dominant along the center line of the
tongue while it vanishes at the boundaries. Numerically we also see
that the $s$ and $d$-wave order parameters form parallel pseudospin
vectors, and the explicit form of the base matrices of these vectors
indicate that the mixed state breaks time reversal symmetry,
i.e., it is an $s+id$ state. This is
consistent with the constraint $A_{{\bf k}}=0$ in Lemma~\ref{lem:T0},
since the constraint requires the two pseudospin vectors to be
parallel unless some other order parameter is non-zero.

  The ferromagnetic (FM) zone that is seen for negative $J$ is just
barely numerically stable close to the phase boundaries.  It is,
however, quite stable deeper inside the FM region.  The numerical
difficulties near the FM phase boundaries can be understood by the
constraint conditions discussed in Lemma~\ref{lem:T0}.  Since, in the
absence of a third order parameter, hopping and FM order cannot exist
simultaneously at the same point in the Brillouin zone, there will be
distinct regions in ${\bf k}$-space.  The FM ordering occurs close to
the Fermi surface, while the hopping expectation value is finite near
the origin in reciprocal space.  The sharp boundary between the two
regions results in discontinuities in the numerical solution.

\subsection{Finite temperature}
We now turn to phase diagrams at non-zero temperature and variable
filling for some particular fixed values of $U$ and $J$.  Estimates
using physical models for the high-$T_c$ materials have suggested that
$U$ should be of the order 5 with $J$ much smaller. The relevance of
such estimates for the present calculation is questionable since they
were made for models of high-$ T_c $ materials that include other
types of interactions such as nearest-neighbor charge repulsion.
Moreover, a large fraction of our Heisenberg term could be considered
as coming from effective second-order corrections to the Hubbard
interaction, which are otherwise neglected in our mean-field approach.
Therefore, we concentrate on parameter values that give interesting
phenomena.

There are two features that we are particularly interested in
exploring.  The first is $d$-wave superconductivity, and the second is
spiral spin-waves.  To keep some connection to high-$T_c$ materials,
we require the model to be antiferromagnetic at half filling, and
therefore $U$ and $J$ should be positive.  In order to study $d$-wave
superconductivity, $U$ should not be too large.  The spiral spin-wave
states, on the other hand, have been observed for the pure Hubbard
model with intermediate $U$, and in this case $J$ should be zero or at
least small.

We start out by investigating the phase diagram for $U=0.5$ and $J=2$
which contains a zone of $d$-wave superconductivity (see
Fig.~\ref{fig:ph_U.5_J2}).  The system is antiferromagnetic close to
half filling.  This state persists up to the N\'{e}el temperature,
where there is a second-order phase transition to the normal state
(NS) that has no broken symmetry.  At low temperatures and moderate
doping, we have a $d$-wave superconductor which is separated from the
antiferromagnet by a first-order phase transition.  For temperatures
higher than the SC critical temperature, there is a first-order phase
transition between the antiferromagnet and the normal state.  This
first-order boundary terminates at higher temperatures at a critical
point, where the transition becomes second order.  No qualitative
differences between this phase diagram and that for $ U = 0 $ is
observed.

Both antiferromagnetic phases and $s$- and $d$-wave superconductors
have been observed in other calculations.\cite{Micnas89}.  Those
studies did not consider the possibility that the phase transition
could be first order, but by carefully comparing the free energies we
have observed this type of transition between the AF and the $d$-wave
SC states at low temperature.  If, as in our model, the total electron
number is fixed, the first-order phase transition results in phase
separation.

By increasing $J$ and $U$, the antiferromagnetic region is
enhanced. The $d$-wave zone grows with increasing $J$, but is also
shifted to larger doping as the antiferromagnetic region expands at
the same time. The overall size of the SC region in the $ T - \mu $
phase diagram is insensitive to a change in $U$.  The $s$-wave
superconducting zone is also enhanced by larger $J$, while it
diminishes if $U$ is increased too much.  It should also be noted that
the AF and the $s$-wave SC regions gain more from an increase in $J$
than the $d$-wave regions do.  This is illustrated in
Fig.~\ref{fig:ph_U0_J4}, where the phase diagram for $U=0$ and $J=4$
is exhibited.  If $J$ is sufficiently large, we expect the $d$-wave to
disappear in favor of the $s$-wave SC and the AF phases.

For large $ U $ and small $ J $, spiral spin-waves appear as shown in
Fig.~\ref{fig:ph_U5_J0} for $J=0$.  The phase diagram is shown in
Fig.~\ref{fig:ph_U5_J0}, where SSW indicates the spiral spin-wave with
the pitch $(\pi-q,\pi)$, which is obtained by applying the twist ${\bf
q}=(1,0)q$ to the AF $(\pi,\pi)$ state.  At low temperature, the SSW
state is separated from the AF state by a first-order phase transition
with a wide coexistence region as a function of density.  For more
elevated temperatures, the separation line becomes second-order and
here the spiral pitch parameter $q$ goes to zero as the phase boundary
is approached.  On the contrary, along the phase transition from the
SSW to the normal state the spiral magnetic order parameter vanishes
in magnitude while $q$ stays finite.  No superconductivity is seen in
Fig.~\ref{fig:ph_U5_J0}.  The $s$-wave superconducting state is
suppressed by the large value of $U$, and the $d$-wave state is not
seen either since $J$ is zero.

The energy gain of the spiral spin-waves is numerically very small, of
the order of a hundredth of the overall condensation energy.  The
introduction of a Heisenberg interaction could therefore have a large
influence.  To investigate this issue, we introduce a small $ J = 0.1
$ while keeping $ U = 5 $, and compare Fig.~\ref{fig:ph_U5_J0} with
Fig.~\ref{fig:ph_U5_J.1}.  What was a second order transition between
the SSW and the normal state for $ J = 0 $ has now become first order.
There is also a temperature range at which the AF is ``reentrant'' as
a function of doping, and where the AF-SSW phase boundary move toward
lower temperatures and becomes first order.  When increasing $J$, this
line of phase transitions rapidly migrate towards lower doping and
very soon the whole spiral spin-wave region is gone.  This explains
why no spiral spin-wave is seen in the phase diagrams for small $U$
and large $J$.

For positive $J$ and sufficiently low densities, the antiferromagnet
and the spiral spin-wave must eventually disappear, leaving room for
the $s$-wave and $d$-wave superconducting states which may persist the
rest of the way to zero filling.  The critical temperature decreases
rapidly with decreasing $J$ and we were unable to confirm this
scenario numerically.

\section{Discussion}
\label{sec:discussion}
The Hubbard model serves as a simple model for high-$T_c$
superconductors.  Although its simple appearance, the model is still
poorly understood and many sophisticated techniques for studying
specific features of the model have been proposed in the literature.
As a guide to this realm of possibilities, it is important to have a
good understanding of all possibilities that a ``simple'' mean-field
analysis can provide.  We have therefore used a generalized
Hartree-Fock-Bogoliubov theory and numerical simulations to compute
phase diagrams for the extended Hubbard model.  All the conventional
order parameters, like $s$ and $d$-wave superconductivity,
charge-density waves, and N\'eel and spiral antiferromagnetic states
states, have been included in one unifying framework, making no a
priori assumptions about the nature of the broken symmetries.  We have
further shown that, in mean-field theory, no new mixed phases arise at
finite doping and temperature in the extended Hubbard model with
positive values of $U$ and $J$. In our investigation, we have seen
the time-reversal symmetry-breaking superconducting phase $s+id$ only
in a narrow region with negative $U$. Close to this region there is also
a region of mixed antiferromagnetism and $d$-wave superconductivity.

Our method allow phase separation to occur, which it also does in
certain regions.  The energy differences that we find between N\'eel
and spiral antiferromagnets is so small, that we do not want to make
any strong statements about whether phase coexistence would survive a
more refined analysis or not.  However, the energy difference between
the AF and the normal state at the first order phase boundary is
substantial. A phase separation between these two states has also been
suggested both from theoretical and experimental
grounds.\cite{Emery93} Another thing to keep in mind is that we
require the total number of electrons to be fixed, while in the
high-$T_c$ materials there are large charge reservoirs surrounding the
2D planes that are perhaps better modeled by a fixed chemical
potential.  Under such circumstances, the phase coexistence may well
be suppressed.

There have been several earlier studies of spiral spin-waves for the
Hubbard model exploiting slave boson and ordinary Hartree-Fock
techniques. Most of these studies have concentrated on zero
temperature, and our corresponding results are consistent with those.
However, we have also extended the analysis to finite temperature.

Another new phenomenon that we have studied is how the spiral
spin-waves are affected by a nearest-neighbor Heisenberg term in the
Hamiltonian.  We have observed that the $(1,0)$ spiral spin-wave phase
is easily destroyed upon the introduction of a positive-$J$ Heisenberg
interaction.  We have mostly concentrated on spin-waves in the
$(1,0)$-direction since we found that the energy differences are very
insensitive to the pitch direction, at least for the pure Hubbard
model.  To make the study complete, other spin directions should also
be studied.  However, it is likely that these small energy differences
are insignificant with respect to the overall crudeness of our
analysis, although one might argue that their relative difference is
to be taken seriously.

\section*{Acknowledgements}
The authors thank the Swedish Natural Research Council (NFR) for
supporting this work.

\appendix
\section{The point group $C_{4v}$}
\label{app:C4}
The point-group symmetry of the 2D square lattice is $C_{4v}$, since
the lattice is invariant under $90^\circ$ rotations around the
$z$-axes and under reflections in the lines $v$ and $v'$ in
Fig.~\ref{fig:C4v_symm_lines}. The group elements of $C_{4v}$ are the
identity ($I$), $90^\circ$ rotations ($C_4$), $180^\circ$ rotations
($C_4^2$), reflections in $v$ ($\sigma_v$) and reflections in $v'$
($\sigma_{v'}$). This group has four 1D irreps ($A_1$, $A_2$, $B_1$,
$B_2$) and one 2D irrep $E$. The character table together with
examples of basis functions for the different irreps are given in
Table~\ref{tab:C4_character}. Our main use of the $C_{4v}$ irreps is
to distinguish between $d$ and $s$-wave superconductivity order
parameters.  The $s$-wave ordering has the full symmetry of the
lattice, {\em i.e.}, it belongs to the $A_1$ representation. The
$d$-wave ordering, on the other hand, is antisymmetric under
reflections in $v'$, and belongs to the $B_1$ representation. If we
would see any $p$-wave states, these would belong to the 2D $E$
representation since these states are antisymmetric under parity
($(x,y)\to(-x,-y)$).

\section{Proofs of the zero-temperature, half-filling theorems}
\label{app:proofsT0}
In this appendix we give the proofs of the theorems concerning the
minimization of the ground-state energy. A bunch of related theorems
have also been derived by Bach {\em et al.}\cite{Bach93} First we
prove the lemma for how the energy minimization problem can be recast
into the problem of minimizing the expectation value of the
Hamiltonian written in terms of $\langle\alpha_{{\bf k}}^m\rangle$.
\begin{duplicate}{Lemma}{\ref{lem:T0}}
The restrictions on the expectation values $\langle\alpha_{{\bf
k}}^m\rangle$ for the variational solutions of
Eq.~(\ref{eqn:generic_ham}) are: $\sum_{m} \langle\alpha_{{\bf
k}}^m\rangle^2 = \frac{1}{2}$, $\langle(\alpha_{00}^0)_{{\bf
k}}\rangle=-\frac{1}{2}$, and $A_{{\bf k}}=0 \; \forall\; {\bf k}$.
\end{duplicate}
Proof.
The space of all possible variational solutions is defined by the
constraint that the Hermitian matrix $\langle G|{\Psi}^{\dagger}_{{\bf
k}}\otimes{\Psi}^{\phantom{\dagger}}_{{\bf k}}|G\rangle$ has the
four-fold degenerate eigenvalues 0 and 1, since it has the same
eigenvalues as $g = \langle G|({\chi}^{\dagger}_{{\bf
k}}\otimes{\chi}^{\phantom{\dagger}}_{{\bf k}})|G\rangle$.  Our aim is
to find the corresponding constraints on the coefficients
$\langle\alpha_{{\bf k}}^m\rangle$ in the expansion
\begin{equation} \label{eq:f-expansion}
\langle G|{\Psi}^{\dagger}_{{\bf k}}\otimes{\Psi}^{\phantom{
\dagger}}_{{\bf k}}|G\rangle = \sum_m \langle\alpha_{{\bf k}}^m\rangle
B_m\,.
\end{equation}
First of all, since $B_0 \equiv B_{00}^0 = -\openone$ is the only
basis matrix with a non-zero trace, one has
$\langle(\alpha_{00}^0)_{{\bf k}}\rangle=-\frac{1}{2}$.  Let us next
define the trace-less matrix $X_{{\bf k}}$
\begin{equation}
\label{eqn:A_def}
X_{{\bf k}} \equiv \langle G|{\Psi}^{\dagger}_{{\bf
 k}}\otimes{\Psi}^{\phantom{\dagger}}_{{\bf k}}|G\rangle -
 {\textstyle\frac{1}{2}}\openone \,.
\end{equation}
This matrix has the fourfold degenerate eigenvalues $\pm \frac{1}{2}$
and the expansion
\begin{equation}
X_{{\bf k}} = \sum_{m \neq 0} \langle\alpha_{{\bf k}}^m\rangle B_m\,.
\end{equation}
Furthermore, $\{X_{{\bf k}},X_{{\bf k}}\}$ has the 8-fold degenerate
eigenvalue $\frac{1}{2}$, meaning that $\{X_{{\bf k}},X_{{\bf k}}\} =
\frac{1}{2}\openone$, so that using Eq.~(\ref{eqn:Bas_orthogonality})
we have
\begin{equation}
\label{eqn:X_sqr}
{\textstyle\frac{1}{2}} \openone = \{X_{{\bf k}},X_{{\bf k}}\} =
 \sum_{m \neq 0} 2\langle\alpha_{{\bf k}}^m\rangle^2 \openone +
 A_{{\bf k}} \,,
\end{equation}
where $A_{{\bf k}}$ is the non-diagonal part defined in
Eq.~(\ref{eqn:Ap_def}).  Since $A_{{\bf k}}$ is orthogonal to
$\openone$, it follows that $A_{{\bf k}}=0$ and that
\begin{equation} \label{eqn:norm}
\sum_m \langle\alpha_{{\bf k}}^m\rangle^2 =
\langle\alpha_{{\bf k}}^0\rangle^2 +
\sum_{m\neq0} \langle\alpha_{{\bf k}}^m\rangle^2 =
(-{\textstyle\frac{1}{2}})^2+{\textstyle\frac{1}{4}} =
{\textstyle\frac{1}{2}} \,.
\end{equation}
Q. E. D.\\
For some specific solutions, this lemma leads to the following theorem
which considerably simplifies the search for solutions.  Here we
consider the larger space of solutions that arises if we disregard all
constraints except the normalization $\sum_m \langle\alpha_{{\bf
k}}^m\rangle^2 = \frac{1}{2}$.  A solution of the new problem that
happens to fulfill {\em all} the constraints, must then be a solution
of the original problem as well.
\begin{duplicate}{Theorem}{\ref{th:T0}}
Consider the Hamiltonian in Eq.~(\ref{eqn:generic_ham}) with fixed
coefficients $a_{l,m}$ and $b_{l,m}$. Define $\cal N$ as the subset of
purely quadratic and irreducible terms, ${\cal N} = \{ m: a_{l,m}=0 \,
\forall\, l\ , b_{l,m}=b_m \delta_{l,l(m)} \}$, where $l(m)$ is a
function which attaches one single $l$ to each $m$.  Assume there
exists an $\tilde{m}\in{\cal N}$, such that ($b_{\tilde{m}}<0$ and
$b_{\tilde{m}}|\beta_{{\bf k}}^{l(\tilde{m})} \beta_{{\bf
k}'}^{l(\tilde{m})}| < b_m|\beta_{{\bf k}}^{l(m)} \beta_{{\bf
k}'}^{l(m)}| \, \forall\, {\bf k},{\bf k}'$, $\forall\, m \in {\cal
N}\setminus\{\tilde{m}\}$).  If ($\langle\alpha_{{\bf k}}^m\rangle=0\,
\forall\, {\bf k},\;\forall\, m\in{\cal N}\setminus\tilde{m}$) is a
sufficient condition for ($A_{{\bf k}}=0 \,\forall\, {\bf k}$), then
the same $\langle\alpha\rangle$'s will be zero also in the minimizing
solution of the constrained problem.
\end{duplicate}
Proof.
The lowest-energy solution should be minimal under any variation of
$\langle\alpha\rangle$'s that fulfills the normalization constraint
(\ref{eqn:norm}).  Since there is only one constraint per ${\bf k}$ in
the simplified problem, it is possible to keep all
$\langle\alpha\rangle$'s fixed except two and still fulfill the
constraint. Suppose that we vary only $\langle\alpha_{{\bf
k}}^{\tilde{m}}\rangle$ and $\langle\alpha_{{\bf k}}^n\rangle$, where
$\tilde{m},n\in{\cal N}$. A variation around a minimizing solution
must then fulfill
\begin{equation}
\left\{\begin{array}{rcl}
{\sum_{{\bf k}'}}'' b_{\tilde{m}}\beta_{{\bf k}'}^{l(\tilde{m})}
\beta_{{\bf k}}^{l(\tilde{m})} \langle\alpha_{{\bf
k}'}^{\tilde{m}}\rangle \delta\langle\alpha_{{\bf
k}}^{\tilde{m}}\rangle & & \\ + b_{n}\beta_{{\bf
k}'}^{l(n)}\beta_{{\bf k}}^{l(n )}\langle\alpha_{{\bf k}'}^n\rangle
\delta\langle\alpha_{{\bf k}}^n\rangle & = & 0 \,,\\
\langle\alpha_{{\bf k}}^{\tilde{m}}\rangle \delta \langle\alpha_{{\bf
k}}^{\tilde{m}}\rangle + \langle\alpha_{{\bf k}}^n\rangle
\delta\langle\alpha_{{\bf k}}^n\rangle & = & 0 \,.
\end{array} \right.
\end{equation}
{}From our assumptions, we have $b_{\tilde{m}}<0$, and to consider
competing solutions one must also have $b_n<0$.  Since there are no
constraints imposed on the signs of $\langle\alpha_{{\bf
k}}^{\tilde{m}}\rangle$ and $\langle\alpha_{{\bf k}}^n\rangle$, it is
obvious that a minimizing solution is obtained by choosing $\rm
sgn(\langle\alpha_{{\bf k}}^{\tilde{m}}\rangle)=sgn(\beta_{{\bf
k}}^{l(\tilde{m})})$ and similarly for $\langle\alpha_{{\bf
k}}^n\rangle$.  Taking these sign considerations into account and
eliminating $\delta \langle\alpha_{{\bf k}}^{\tilde{m}}\rangle$ and
$\delta \langle\alpha_{{\bf k}}^n\rangle$ yields
\begin{eqnarray}
&&{\sum_{l,{\bf k}'}}'' (b_{\tilde{m}} |\beta_{{\bf k}}^{l(\tilde{m})}
\beta_{{\bf k}'}^{l(\tilde{m})}| |\langle\alpha_{{\bf
k}'}^{\tilde{m}}\rangle \langle\alpha_{{\bf k}}^n\rangle| \nonumber \\
&& - b_{n} |\beta_{{\bf k}}^{l(n)}\beta_{{\bf k}'}^{l(n)}|
|\langle\alpha_{{\bf k}'}^n\rangle\langle\alpha_{{\bf
k}}^{\tilde{m}}\rangle|) = 0\,.
\end{eqnarray}
Summing this equation over ${\bf k}$ gives
\begin{equation}
{\sum_{{\bf k},{\bf k}'}}'' (b_{\tilde{m}} |\beta_{{\bf
k}}^{l(\tilde{m})} \beta_{{\bf k}'}^{l(\tilde{m})}| - b_n |\beta_{{\bf
k}}^{l(n)}\beta_{{\bf k}'}^{l(n)}| ) |\alpha_{{\bf
k}'}^{\tilde{m}}\alpha_{{\bf k}}^n| = 0\,,
\end{equation}
and since from our assumption, $b_{\tilde{m}} |\beta_{{\bf
k}}^{l(\tilde{m})} \beta_{{\bf k}'}^{l(\tilde{m})}| < b_n |\beta_{{\bf
k}}^{l(n)}\beta_{{\bf k}'}^{l(n)}|\, \forall \,{\bf k}, {\bf k}'$, the
solution must be either $\alpha_{{\bf k}}^{\tilde{m}}=0\,\forall\,
{\bf k}$ or $\alpha_{{\bf k}}^n=0\,\forall\,{\bf k}$.  Of these two,
$\alpha_{{\bf k}}^n=0$ is obviously the solution of lowest
energy. This argument is then applied to all purely quadratic terms
$n\in{\cal N}\setminus\tilde{m}$. \\
Q. E. D.\\

\renewcommand{\baselinestretch}{1.1}

\renewcommand{\baselinestretch}{0.9}
\input{epsf}
\newpage
\newlength{\figurewidth}
\newlength{\smallfigwidth}
\setlength{\figurewidth}{10cm}
\setlength{\smallfigwidth}{8cm}
\narrowtext

\begin{figure}

\centerline{
\epsfxsize=\smallfigwidth
\epsfbox{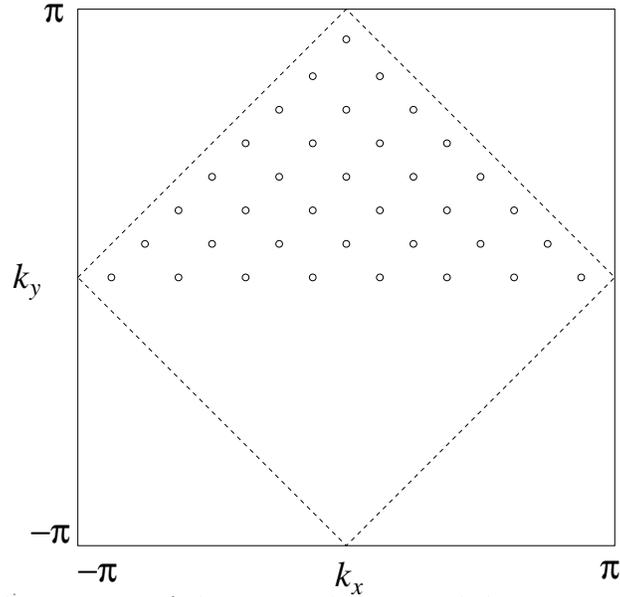}
}

\caption{The first Brillouin zone of the square lattice with lattice
constant 1. The line $\epsilon_{{\bf k}}=0$ is indicated as well as
the ${\bf k}$-points in the reduced zone that are used in a numerical
simulation with 32 ${\bf k}$-points
(Sec.~\protect\ref{sec:numerical}).  The contributions from the points
on the $k_x$-axes are scaled down by a factor of two since they would
otherwise be overcounted.}
\label{fig:1BZ}
\end{figure}

\begin{figure}

\centerline{
\epsfxsize=\figurewidth
\epsfbox{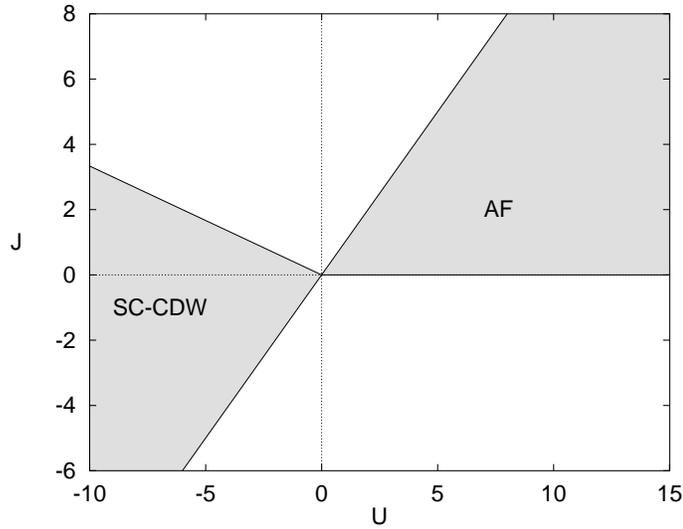}
}

\caption{The low energy states of the extended Hubbard model according
to Corollary~\protect\ref{corr:T0} is shown in grey. The white areas
are indeterminate from the corollary, and the phases here must be
computed numerically.}
\label{fig:ph_corr}
\end{figure}

\begin{figure}

\centerline{
\epsfxsize=\figurewidth
\epsfbox{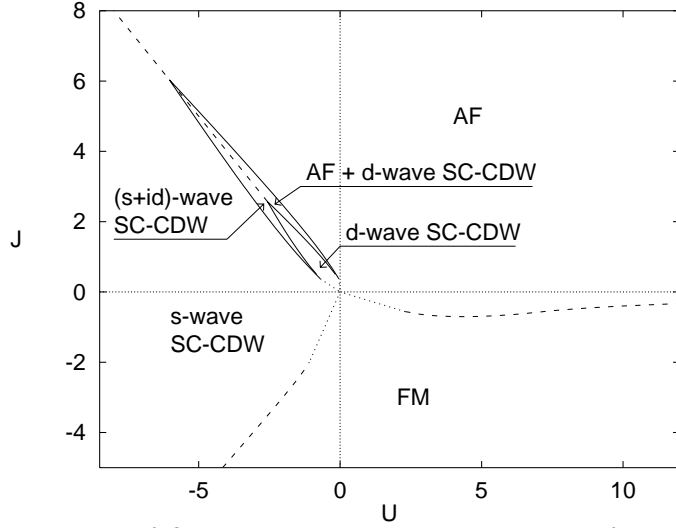}
}

\caption{Phase diagram at half filling and zero temperature as a
function of $J$ and $U$.  Second-order phase boundaries are drawn as
full lines, while first order phase boundaries are drawn as dashed
lines. The dotted lines are extrapolations of the numerically derived
full (dashed)-line phase boundaries.}
\label{fig:pd_T0_mu0}
\end{figure}

\begin{figure}

\centerline{
\epsfxsize=\figurewidth
\epsfbox{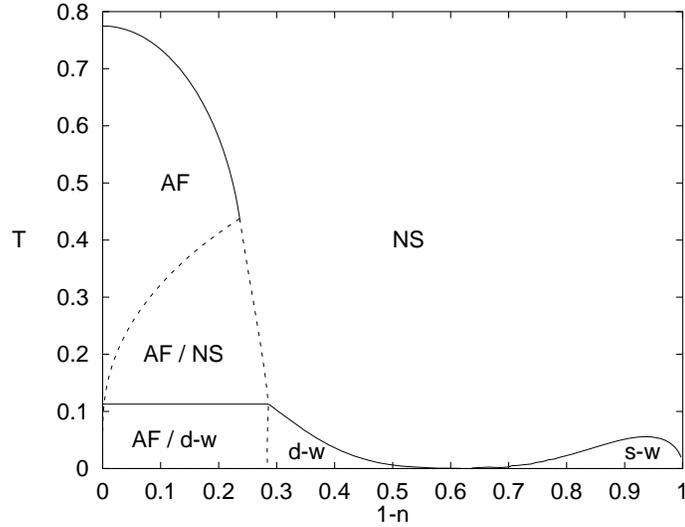}
}

\caption{Phase diagram for $U=0.5$, $J=2$. Second-order phase
transitions are drawn as full lines, while the boundaries of regions
with phase separation are drawn as dashed lines. The regions of phase
separation are denoted A / B where A and B are the two coexisting
phases.  Regions of $s$-wave and $d$-wave SC are indicated by s-w and
d-w, respectively.}
\label{fig:ph_U.5_J2}
\end{figure}

\begin{figure}

\centerline{
\epsfxsize=\figurewidth
\epsfbox{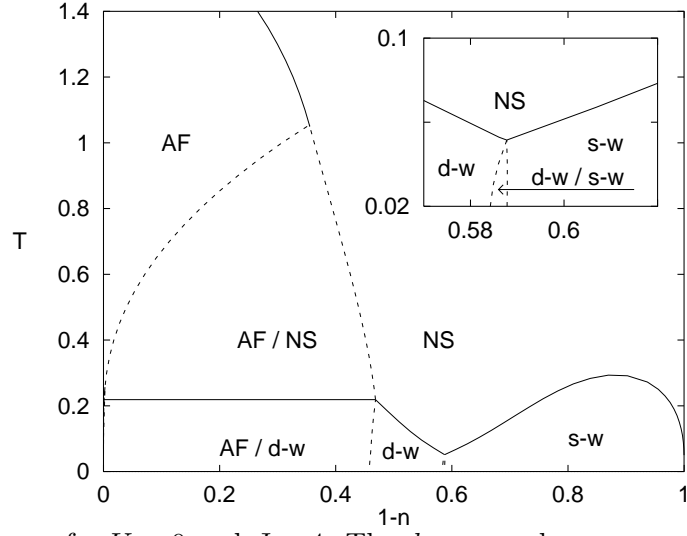}
}

\caption{Phase diagram for $U=0$ and $J=4$. The $d$-wave and
$s$-wave superconducting regions are separated by a first order phase
transition, with a very narrow coexistence region. The inset shows a
magnification of the coexistence region.}
\label{fig:ph_U0_J4}
\end{figure}

\begin{figure}

\centerline{
\epsfxsize=\figurewidth
\epsfbox{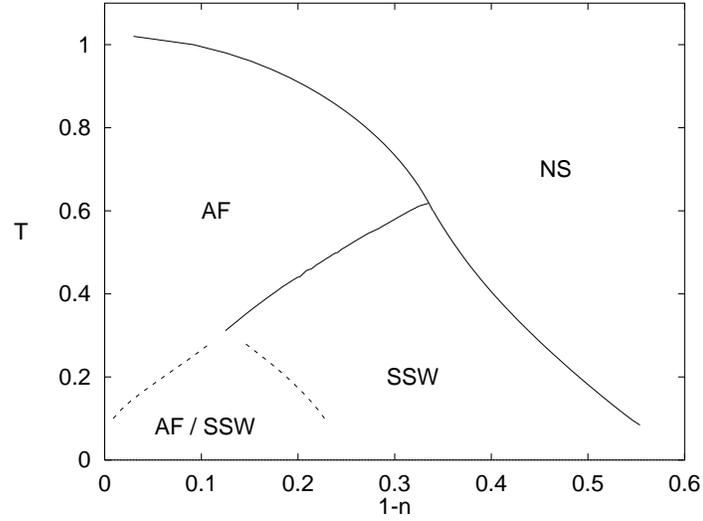}
}

\caption{Phase diagram for $U=5$ and $J=0$. The spiral spin-wave
with pitch $(\pi-q,\pi)$ is denoted SSW.}
\label{fig:ph_U5_J0}
\end{figure}

\begin{figure}

\centerline{
\epsfxsize=\figurewidth
\epsfbox{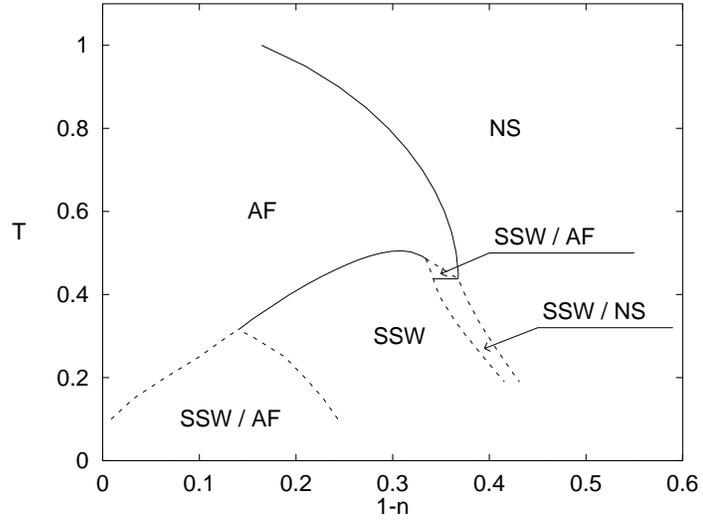}
}

\caption{Phase diagram for  $U=5$, $J=0.1$. The SSW has pitch
$(\pi-q,\pi)$.}
\label{fig:ph_U5_J.1}
\end{figure}

\begin{figure}

\centerline{
\epsfxsize=\smallfigwidth
\epsfbox{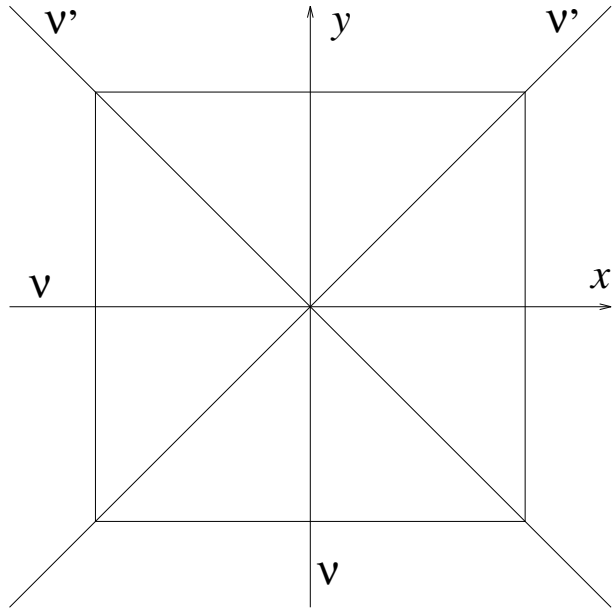}
}

\caption{The symmetry axes $v$ and $v'$ of the square lattice.}
\label{fig:C4v_symm_lines}
\end{figure}

\newpage
\narrowtext
\begin{table}
\caption{Construction of the $8\times 8 $ $\beta_A$ matrices from the
$4\times 4$ Dirac matrices expressed in terms of Pauli matrices. The
index $j$ runs from 1 to 3. The notation $\gamma_j^*$ denotes complex
conjugate ({\em not adjoint}).}
\label{tab:gamma_mat}
\renewcommand{\arraystretch}{0.8}
\begin{tabular}{c|l}
&\\[-6pt] Pauli & $\left(\sigma_1,\sigma_2,\sigma_3\right) =
\left(\left(\begin{array}{cc} 0 & 1 \\ 1 & 0 \end{array}\right),
\left(\begin{array}{cc} 0 & -i \\ i & 0 \end{array}\right),
\left(\begin{array}{cc} 1 & 0 \\ 0 & -1 \end{array}\right)\right)
$\\[12pt] $\begin{array}{l} 4\times 4 \\ \mbox{Dirac}\end{array}$ &
$\gamma_0 = \left(\begin{array}{cc}\openone & 0 \\ 0 &
-\openone\end{array}\right)$ \ \ \ \ $\gamma_j =
\left(\begin{array}{cc} 0 & \sigma_j \\ -\sigma_j & 0
\end{array}\right)  $  \\[12pt]
$\begin{array}{l} 8\times 8 \\ \beta_A\end{array}$ & $\beta_0 =
\left(\begin{array}{cc} \gamma_0 & 0 \\ 0 & -\gamma_0
\end{array}\right)$ \ \ \ \ $\beta_j = \left(\begin{array}{cc}
\gamma_j & 0 \\ 0 & \gamma_j^* \end{array}\right)$ \\[12pt] &
$\beta_{j+3} = iZ\beta_0\beta_jZ$
\end{tabular}
\end{table}

\begin{table}
\caption{The character table of the point group $C_{4v}$ together with
examples of basis functions for the different irreducible
representations.}

\begin{tabular}{l|rrrrr@{\hspace{50pt}}l}
 & $I$ & $C_4^2$ & $C_4$ & $\sigma_v$ & $\sigma_{v'}$ & Examples of
functions \\ \tableline $A_1$ & 1 & 1 & 1 & 1 & 1 & 1,
$\cos{k_x}+\cos{k_y}$ \\ $A_2$ & 1 & 1 & 1 & -1 & -1 &
$\sin{2k_x}\sin{k_y}-\sin{2k_y}\sin{k_x}$ \\ $B_1$ & 1 & 1 & -1 & 1 &
-1 & $\cos{k_x}-\cos{k_y}$ \\ $B_2$ & 1 & 1 & -1 & -1 & 1 &
$\sin{k_x}\sin{k_y}$ \\ $E$ & 2 & -2 & 0 & 0 & 0 & $\{\sin{k_x}$,
$\sin{k_y}\}$
\end{tabular}
\label{tab:C4_character}
\end{table}

\end{document}